\documentclass[journal]{IEEEtran}
\usepackage{multirow}
\usepackage{graphicx}
\usepackage{subfigure,tabularx}
\usepackage{amsfonts,amsmath,color,amssymb,amsxtra,graphicx,floatflt}
\usepackage{pifont}
\usepackage{enumerate}

\usepackage{fancyhdr}

\usepackage{hyperref}
\hypersetup{
    colorlinks=true,
    linkcolor=black,
    filecolor=magenta,      
    urlcolor=cyan,
    citecolor=black,
}
 
\fancypagestyle{IEEE_Green_open_access_footer}{
  \fancyhf{}
  \fancyhead[C]{\footnotesize \copyright2019 IEEE. Personal use of this material is permitted. Permission from IEEE must be obtained for all other uses, in any current or future media, including reprinting/republishing this material for advertising or promotional purposes, creating new collective works, for resale or redistribution to servers or lists, or reuse of any copyrighted component of this work in other works.     DOI: 10.1109/IOTM.0001.1800022}     

}

\begin{document}
%
\title{A Standard-based Open Source IoT Platform: FIWARE}

\author{Flavio Cirillo\IEEEauthorrefmark{1}\IEEEauthorrefmark{2}, G\"urkan Solmaz\IEEEauthorrefmark{1},  Everton Lu\'is Berz\IEEEauthorrefmark{1}, Martin Bauer\IEEEauthorrefmark{1}, Bin Cheng\IEEEauthorrefmark{1}, Ernoe Kovacs\IEEEauthorrefmark{1}\\
\IEEEauthorrefmark{1} NEC Laboratories Europe, Heidelberg, Germany\\
\IEEEauthorrefmark{2} University of Naples Federico II, Naples, Italy\\
E-mails: \{name.surname\}@neclab.eu
}

\maketitle
\thispagestyle{IEEE_Green_open_access_footer}

\begin{abstract}
The ever-increasing acceleration of technology evolution in all fields is rapidly changing the architectures of data-driven systems towards the Internet-of-Things concept. Many general and specific-purpose IoT platforms are already available. 

This article introduces the capabilities of the {\em FIWARE} framework that is transitioning from a research to a commercial level. We base our exposition on the analysis of three real-world use cases (global IoT market, analytics in smart cities, and IoT augmented autonomous driving) and their requirements that are addressed with the usage of FIWARE. We highlight the lessons learnt during the design, implementation and deployment phases for each of the use cases and their critical issues. Finally we give two examples showing that FIWARE still maintains openness to innovation: semantics and privacy.

\end{abstract}

\IEEEpeerreviewmaketitle

\section{Introduction}

The evolution in all fields of technology is accelerating the shift towards the Internet-of-Things (IoT) vision in any area where data are produced and analyzed. 

The market of IoT platforms is comprised of a huge number of solutions. From a business perspective, we can cluster them in two classes of platforms: commercial and open source. The commercial platforms are more stable and more appealing for industrial and business-oriented scenarios due to the contractual support of the providers. The open-source platforms are usually implementing standards and maintained by communities of researchers and innovators, enhancing them with cutting-edge trends, driven by use case and practical requirements. Here we discover the transition of FIWARE\footnote{\url{https://www.fiware.org/}} from research to commercial.

\subsection{FIWARE and its evolution}
The FIWARE framework is supported by the European Commission since almost a decade, through funding several projects either for developing and enhancing it (e.g., FI-WARE, FI-Core, FI-Next, SmartSDK) or for using it in pilots (e.g., City Platform as a Service - CPaaS.io\footnote{\url{https://cpaas.bfh.ch/}}, frontierCities, Fiware4Water). Lately, the FIWARE framework is going through a transition phase towards business readiness. This is demonstrated by the European large scale pilots projects (each with 15-20 Million Euro of funds), such as SynchroniCity\footnote{\url{https://synchronicity-iot.eu/}}, Autopilot\footnote{\url{https://autopilot-project.eu/}}, Internet of Food and Farm - IoF2020, and Activage focusing on smart cities, autonomous driving, agriculture \& food and e-health, respectively.

\section{Public governance and growth: global IoT market}
\label{sec:usecase-synchronicity}
Public governance is attentive to new technologies that have a strong potential for economic growth, public safety, and citizens' well-being. IoT is certainly promising all these benefits. This has triggered in the past decade a widespread adoption of a plethora of IoT solutions in many urban scenarios. However, closed commercial APIs and implementations hamper an open market development (vendor lock-in). For this reason, an open approach is preferred by public institutions, as commonly agreed, for instance, by the 140+ cities of the Open \& Agile Smart Cities (OASC) network\footnote{\url{http://oascities.org/}}. Horizontal harmonization is provided locally in cities such as Milan\footnote{\url{http://www.milanosmartcity.org/joomla/}} and Helsinki\footnote{\url{https://www.helsinkismart.fi/}} achieving good results but remaining regionally isolated, thus resulting in city lock-in. Both vendor lock-in and city lock-in discourage small and medium-sized enterprises (SMEs). 

With this rationale, SynchroniCity project proposes to synchronize existing smart city solutions by overcoming their misalignments. More specifically, the synchronization targets enabling five Minimal Interoperability Mechanisms (MIMs):
\begin{enumerate}
    \item Context Management
    \item Data Models
    \item Ecosystem Transaction Management (''Marketplace'')
    \item Security
    \item Storage
\end{enumerate}
In 2019 the first three MIMs have been officially adopted by the OASC consortium~\cite{oasc_mim}.

The SynchroniCity project demonstrates the feasibility of such an approach by harmonizing eight cities (Antwerp, Carouge, Eindhoven, Helsinki, Manchester, Milan, Porto, and Santander).
The final goal is the co-design of IoT services~\cite{Cirillo-AtomicServices} and IoT applications, and the establishment of a shared market fostering economic growth. 

\begin{figure}[t!]
\centering
\includegraphics[width=\columnwidth,trim={8cm 7.5cm 23cm 8cm},clip]{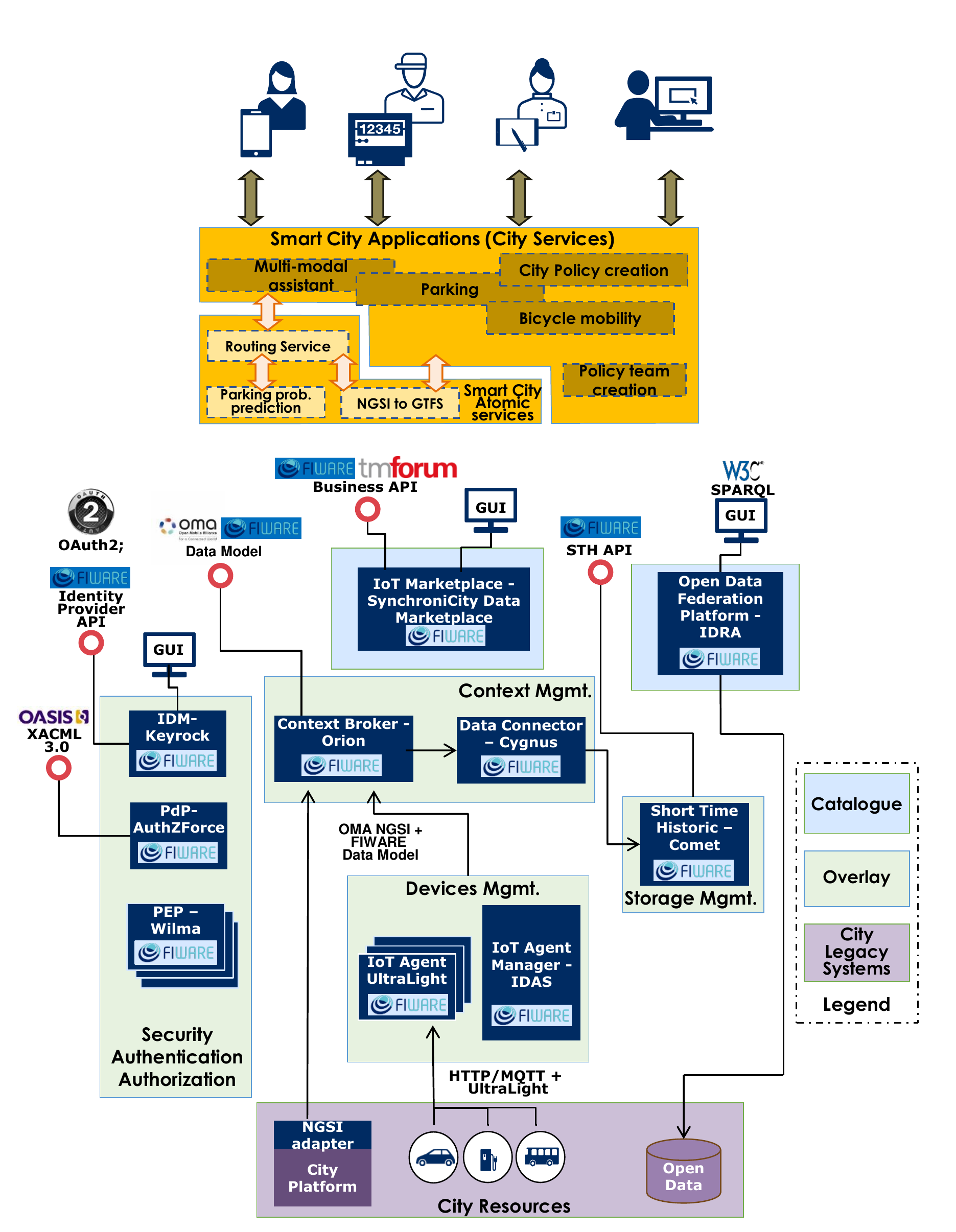}
\caption{Federation of pilot sites through the FIWARE IoT platform.}
\label{Fig:SynchroniCity}
\end{figure}

\subsection{Legacy IoT platforms are not neglected}

In today's smart cities it is common to have platforms that are the outcome of past projects and pilots of city governments. Often these solutions are ad-hoc and only used for the local smart city ecosystem and market. City governments are not keen on throwing away the obtained achievements, but they are interested in expanding the IoT business. Thus, instead of re-designing systems and replacing the already operational ones, in SynchroniCity a generic overlay platform over existing infrastructures is realized. This overlay finds in the FIWARE framework many building blocks to implement the five foreseen MIMs (see Fig.~\ref{Fig:SynchroniCity}). Starting from the available city resources (the purple box in Fig.~\ref{Fig:SynchroniCity}) data is integrated and exposed in several manners: sensor streams, open data, and historical time series.

\subsection{Harmonizing smart cities data}

The sensor streams are integrated with different approaches depending on the requirements and constraints given by the different city platforms. 
For most of the involved cities, ad-hoc integration modules are implemented and deployed. These obtain data from the running platforms and translate them to the Next Generation Service Interface (NGSI)~\cite{OMA-NGSI-Context} standard data format.
Only in a few cases, such as Santander and Porto, sensor data are exposed through typical IoT interfaces, such as MQTT and HTTP. For native IoT device interfaces, the FIWARE framework offers IoT Agents (see Table~\ref{tab:iotagents}) that translate device interfaces to NGSI. 
Having all data formatted using the same data structure is not enough to ensure harmonization since data might be modeled very differently by IoT developers. For instance, a location might be referred to as "position", "geolocation" or a term in another language. Therefore, the used adapters, both ad-hoc modules and IoT Agents, generate context data following the FIWARE/OASC data models\footnote{\url{https://www.fiware.org/developers/data-models/}}. These data models are defined by the FIWARE community and adopted by OASC, and they harmonize the description of data for several application areas such as parks and gardens, points of interest, parking, waste management, transportation, and weather.

\begin{table}[]
\caption{Device interfaces supported by FIWARE}
\centering
\begin{tabular}{|c|c|}
\hline
\textbf{IoT Devices} & HTTP, MQTT, AMQP\\ \hline
\textbf{\begin{tabular}[c]{@{}c@{}}Low-power \\wireless sensors\end{tabular}} & LoRaWAN, SigFox  \\ \hline
\textbf{Smart Industry} & OPC-UA, ROS2 \\ \hline
\end{tabular}
\label{tab:iotagents}
\end{table}

The homogeneously modeled data is then handled by the context management layer that exposes context with a standard interface, which is NGSI, fulfilling the Context Management MIM. NGSI specifies both a context management interface with an HTTP binding and a context data format using JSON.
SynchroniCity adopted the Orion Context Broker (CB) as the context manager implementation, which holds the latest attribute values for each entity (or ''thing'') and exposes NGSI query and subscription methods. Orion is conceived to work with high-frequency messages and to respond to queries with minimal latency. For stream-based applications, Orion offers an NGSI subscription interface, notifying with atomic notifications as soon as a matching attribute of a matching entity becomes available. A sensor generating a stream of data with small intervals between observations, such as an accelerometer, might create a flood of notifications in the network. Thus, a ``throttling'' parameter can be set in the subscription. This regulates data notification streams, instructing Orion to issue two notifications for the same subscription apart in time for at least the throttling period. The drawback is that it might result in data loss in case more than one value for a matching attribute is generated within the throttling period. 
The missing value is not an option for some applications. For this reason, SynchroniCity adopted the Comet Short Time Historic (STH)\footnote{\url{https://github.com/telefonicaid/fiware-sth-comet}} component. The Cygnus connector receives the NGSI data stream from Orion and forwards it to a persistent data sink, such as STH Comet (in the SynchroniCity case), but also to other commonly used data storage systems such as MongoDB, Hadoop Distribute File Systems (HDFS) for big data processing, or CKAN for Open Data publication. STH and Cygnus together address the Storage MIM.

\subsection{IoT Marketplace}
Often, in smart cities, there are already many services that produce and use data for their purposes, such as urban facilities, public transportation, tourism operators. Unfortunately, these IoT providers have no interest in spending effort on sharing their data if there is no payback for it. On the other side, companies that might want to create smart city services need data and they are keen on paying a fee for datasets otherwise impossible to get. 
Here the necessity of having an IoT marketplace arises. Available data needs to be cataloged: a) as valuable assets in case of private companies, b) as open data in case of public institutions. 

For the first scenario, a gap is identified, and, therefore, the project worked closely with the FIWARE community to close it. FIWARE officially adopted the TM Forum business API standard which is implemented within the SynchroniCity project. 
This software component, named SynchroniCity IoT Data Marketplace\footnote{\url{https://iot-data-marketplace.com}}, exposes a catalog of available data (either in Orion or in STH Comet) describing endpoints to access them, license, and price. 
Users can buy data items and the marketplace sets access rules in the security layer to allow the data exchange. 
To smooth the usage of the rather complex business API, scripts are available to automatically integrate the FIWARE data management by crawling Orion and publish the found data items in the marketplace.

In the case of open data, smart city governments usually have already published datasets, most of the time on their ad-hoc platforms or simply on their institutional website. 
That is not handy for a data consumer that is completely unaware of the city governance structure and, therefore, of the different institutional web pages. The situation is even worse when the open data website is only available in the local language. This situation is addressed by the Open Data Federation Platform IDRA\footnote{\url{https://opendata.synchronicity-iot.eu/}} that was developed in the context of the FESTIVAL EU project that faced a similar scenario. Available datasets are federated in IDRA either via typical open dataset interfaces (such as CKAN) or simply via the web scraping functionality that crawls the generic website for datasets. All the federated datasets are then exposed by a graphical user interface. 

\subsection{Security}
Identity management and access control requirements of SynchroniCity are not specific to smart cities and IoT. Therefore, common standards are used, such as OAuth2 protocol for authorization, and Oasis eXtensible Access Control Markup Language (XACML) 3.0 protocol for access control. The FIWARE framework already offers software components for both, namely KeyRock and AuthZForce respectively, that are already well integrated with the other FIWARE components (such as the IoT marketplace).

\subsection{Lessons learned for the creation of a Smart Cities global market}
Forgoing from a local to a global market, SynchroniCity tackles three aspects: the IoT service interface, the data models, and an IoT marketplace where data can be bought and sold. 

The first barrier faced by SynchroniCity is that putting aside running legacy systems is not a solution for city governments, even if this enables a business ecosystem and, perhaps, it boosts local IoT economic growth. Thus, an overlay on top of the running IoT infrastructure is designed and deployed in every pilot city. 
The overlay harmonizes live sensors data and open data. FIWARE offers off-the-shelf solutions in case of data exposed through common IoT interfaces (see Table~\ref{tab:iotagents}), and for open data (i.e., IDRA). In the case of data exposed through proprietary platforms, the burden of creating ad-hoc adapters cannot be avoided.

A technical issue occurred when data streams from sensors with very frequent observations are not optimally handled by the Orion CB component. With the latter, either too many notifications flood the network, or data may be lost. The solution for SynchroniCity is to use a time-series database as a data sink. A different approach to this problem is the usage of the Aeron IoT Broker component, as in the CPaaS.io scenario (see below).

The fact that FIWARE is supported by a community directly helps in making such a framework useful for IoT scenarios since it is driven by practical problems. Indeed, SynchroniCity could leverage the outcome of other projects to solve open data integration. SynchroniCity itself contributes to the IoT marketplace, releasing it as free to use\footnote{https://github.com/capossele/SynchroniCityDataMarketplace}, thus, closing a gap in FIWARE.
\section{Data analytics on IoT federation: a smart city scenario}

IoT services require IoT data. As seen above, in a smart city data may come from different public providers, such as public transportation or traffic management, and it may be centrally handled by the city governance. But what happens when data comes from private entities, such as a company, or even private citizens? Such private providers like to keep their own IoT infrastructure and their data, and not give data away to a centralized unknown platform. Providers are willing to share their datasets~\cite{PWC}, if they are licensed and protected by access control, or even to earn money considering data as valuable assets. 
Such a scenario of fragmented IoT platforms is a nightmare for IoT service providers since they need to look for IoT providers and make a great effort on integrating heterogeneous data sources. Furthermore, an IoT service is typically composed of multiple data analytics routines, each exploiting a different set of IoT data and depending on each other.
And what would happen if the IoT service provider wants to port its service in another environment? Simply additional effort on data providers discovery, data integration, and analytics orchestration.

\subsection{City Platform as a Service}

The CPaaS.io~\cite{Koshizuka-CPaaSio} EU-Japan project (City Platform as a Service-Integrated and Open) faced these problems in a smart city scenario. To overcome such a situation, CPaaS.io defined the following fundamental requirements to be addressed:
\begin{enumerate}[i)]
    \item allow easy integration of data sources into the platform (e.g., sensors operated by private citizens or established providers), 
    \item offer services and federation capabilities among IoT platform instances,
    \item support the deployment in many cities with distinct requirements (flexibility and elasticity),
    \item self-orchestration of data analytics processing routines, each part of the same IoT service, among multiple IoT data providers,
    \item provide security mechanisms for the desired privacy and data protection,
    \item demonstrate that the use cases developed in one city can be transferred to other cities.
\end{enumerate}

To test the feasibility and effectiveness of the solution, CPaaS.io targets three smart city use cases spread among Europe (Amsterdam) and Japan (Sapporo, Yokosuka, Tokyo), such as water management, public event management, and public transportation.

The CPaaS.io platform exploits different technology to address the identified requirements, such as IoT platform federation, IoT data access control, security \& privacy and data analytics tasks orchestration. 
Also, in this case, the FIWARE framework provides support on the European side, while the u2 platform~\cite{u2Japan} is used on the Japanese side.
The solution of adopting two heterogeneous platforms validates the viability of a loosely coupled federation in the real world~\cite{Koshizuka-CPaaSio}.

The European platform is FIWARE-based where FIWARE components are deployed and, in some cases, enhanced to provide the basis for a smart city data infrastructure. 
Table~\ref{tab:cpaasio} presents a list of the FIWARE framework components in use by the CPaaS.io project and the mapping of each component onto the CPaaS.io functional architecture.

\begin{table}
\centering
\caption{FIWARE-based components present in the CPaaS.io project}
\begin{tabular}{ll} 
\hline\hline
CPaaS.io architecture layer                   & Component               \\ 
\hline\hline
\multirow{2}{*}{Security and Privacy}  & KeyRock                 \\
                                              & PEP-Proxy               \\ 
\hline
\begin{tabular}{@{}l@{}}Data Analytics Routines\\Management and Operation\end{tabular}            & FogFlow                 \\ 
\hline
\multirow{3}{*}{Semantic Data \& Integration} & IoT Knowledge Server    \\
                                              & NGSI to RDF Mapper      \\
                                              & STH Comet               \\ 
\hline
\multirow{2}{*}{Virtual Entity}         & IoT Discovery           \\
                                              & Context Broker          \\ 
\hline
IoT Data and Ingestion                  & IoT Broker              \\ 
\hline
IoT Resource                            & IoT Agent: LoRaWAN to NGSI  \\
\hline\hline
\end{tabular}
\label{tab:cpaasio}
\end{table}

\subsection{Urban water management scenario}
As a relevant application scenario, Waterproof Amsterdam uses the CPaaS.io platform for water management. In the urban context, periods of drought and sudden heavy flashes of rainfall occur more and more often due to the urbanization trend and global warming. In both weather conditions, smart water management is required to ensure water availability, and to handle high pressure in the sewerage infrastructure to prevent street floods.
The solution to this problem is water supply peak shaving. This is done by using smart and integrated rain buffers, such as rain barrels, retention rooftops, and retention buffers. These smart devices are centrally controlled by the IoT water management service that makes use of weather information, sewerage capacity, and environment data to calculate the optimal water filling degree for each buffer. 

The Waterproof application scenario necessitates several functional layers:
\begin{itemize}
    \item \textit{IoT Resources layer} for handling the communication with devices and sensors.
    \item \textit{IoT Data and Ingestion layer} to persist and index, and to expose data to IoT data consumers.
    \item \textit{Virtual Entity layer} that holds representations in the virtual world of the observed things.
    \item \textit{Data Analytics Routines Management and Operation layer} to orchestrate the different data processing tasks for computing the operations for the smart devices.
\end{itemize}

\begin{figure}[t!]
\centering
\includegraphics[width=0.6\columnwidth,trim={0cm 9.47cm 19.46cm 0cm},clip]{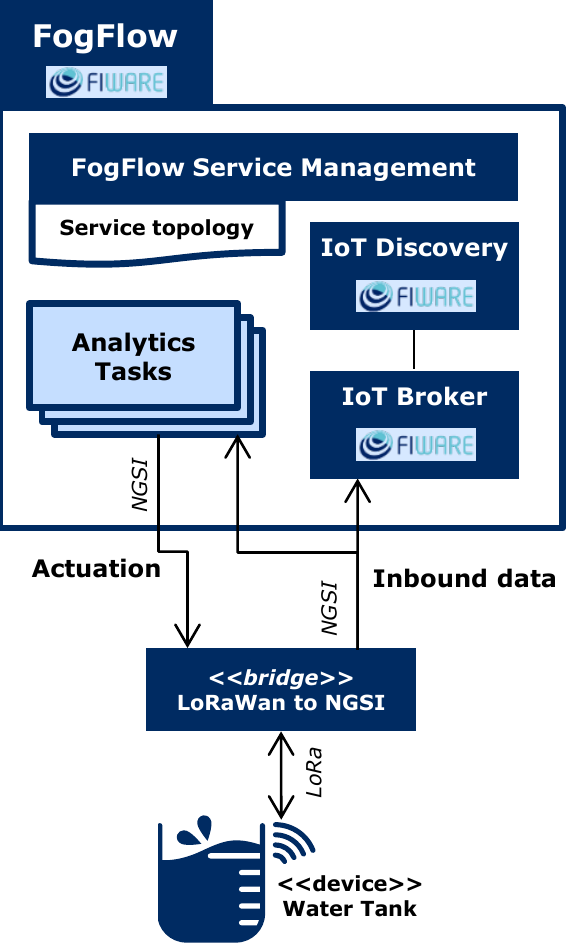}
\caption{Waterproof architecture in the CPaaS.io platform}
\label{Fig:Waterproof}
\end{figure}

The overall system architecture of the Waterproof solution is presented in 
Figure~\ref{Fig:Waterproof}. Similar to the SynchroniCity case, the IoT resources are handled through IoT Agents, but in this case, using the Long Range Wide Area Network (LoRaWAN) protocol. Even though a LoRaWAN IoT Agent is already available in the FIWARE framework, a new one has been implemented in order to optimize it for the specific Waterproof use case, starting from the FIWARE IoT Agent library\footnote{\url{https://iotagent-node-lib.readthedocs.io}} for accelerating the software development. For handling data, instead of the Orion Context Broker, as in SynchroniCity, the Aeron IoT Broker is used, which has additional features such as federation readiness both for query and subscription~\cite{Cirillo-LIoTS}, and throttling-based aggregation of notifications. The latter permits to receive data notifications not more often than the throttling but encompassing all sensor observations pushed within a period through aggregation such as to avoid data loss. The aggregation can be appending all the data values in a set, or applying an actual function (e.g., averaging). The drawback is a bit higher latency compared to Orion, which is acceptable in the Waterproof use case as it is a smaller IoT scenario compared to a complete city in the case of SynchroniCity. Afterward, all the data are stored in the STH historical component, which is extended to handle also the historical storage of metadata.

The virtual world representation is kept within the FIWARE IoT Discovery which acts as a registry of the available data and resources with associated metadata. This layer abstracts real-world objects (things) from the observation points (sensors). The IoT Discovery allows us to query and subscribe for resources by requesting the relevant context, such as a street name.

The Waterproof data analytics is broken down into atomic analytics tasks that are linked in a chain or a generic topology. This brings the advantage of flexibility. With such an approach, the plugging of the right data streams to the tasks' inputs is not hardcoded, and, therefore, an external process can automatize it. For this purpose, CPaaS.io uses the FIWARE FogFlow~\cite{FogFlow} framework. The latter leverages the virtual world representation to discover the needed resources and plugs them to task inputs. The resources can be physical (e.g.,  a sensor stream), or generated, such as intermediate results of one of the analytics tasks (e.g., geographical data aggregation). With such an approach, porting the service topology into another environment is smooth, since FogFlow works at the higher abstraction level of the virtual world.
When porting the Waterproof application, the only configuration needed is the description of the available computing nodes.
In this application, analytics tasks are responsible for identifying heavy precipitations and for computing how to set up the water network to avoid damage. The analytics service subscribes to inbound data streams to constantly monitor the state of the water buffer and trigger actions when needed. 

FogFlow is conceived to orchestrate tasks also among cloud and edge~\cite{FogFlow}, which is a feature to be used in the future for exploiting the computing resources of the deployed devices.

\subsection{Lessons learned from smart city services}

Federating IoT systems is often not an option~\cite{Cirillo-LIoTS}. Therefore, CPaaS.io sets federation as a target aiming at integrating private and heterogeneous IoT systems towards a global Internet-of-Things. Luckily, the NGSI protocol has been designed to support federation, and the FIWARE framework already offers implemented solutions with the Aeron IoT Broker and NEConfMan IoT Discovery components~\cite{Cirillo-LIoTS}.

The water management use case helped to identify other technical gaps not foreseen beforehand. For instance, as the LoRaWAN to NGSI bridge component requires high throughput and low latency, a new lightweight component has been developed by The Things Network\footnote{\url{https://www.thethingsnetwork.org}} to address those performance needs.
Another gap was the support of metadata in time series storage that is addressed by extending the STH component. In NGSI, metadata means additional data about attribute values. 
For instance, in the Waterproof application, the time of measurement and measurement unit are metadata used by the water management system and the metadata need to be historically persisted together with attribute values.
The implementation of the new LoRaWAN IoT Agents and the extension of the STH component have been made as part of the CPaaS.io project thanks to the open-source nature of the FIWARE framework. In fact, in the first case, the available library has been used in the implementation and in the second case the source code of the component has been extended.

\section{Research and innovation: an automated driving scenario}

In CPaaS.io and SynchroniCity, we have seen requirements for smart cities. But how can we handle extreme scenarios where applications need very low latency, high computation power, ubiquitous computing, and different administrative IoT domains? This is the boundary situation of the smart mobility domain which is the main focus of the AUTOPILOT project.

AUTOPILOT aims to leverage IoT technologies for enhancing automated driving. The smart mobility use cases include applications of car sharing, car re-balancing, autonomous platooning, and automated valet parking. One key aspect of the future smart mobility scene is providing interoperability between various components which are in-vehicle (e.g., autonomous cars), on-site (e.g., roadside units - RSUs), or in the cloud (e.g., traffic operation center). Hence, the project aims to support the flow of IoT information from the vehicle to the cloud and back to the vehicle passing through on-site units and network infrastructures. Besides, such applications need to be able to work when a car passes from one city to another, and, therefore, it passes road administration boundaries typically handled by different institutions.

\subsection{Federation of large scale pilots}

\begin{figure}[t!]
\centering
\includegraphics[width=\columnwidth,trim={0.5cm 1cm 1cm 3cm},clip]{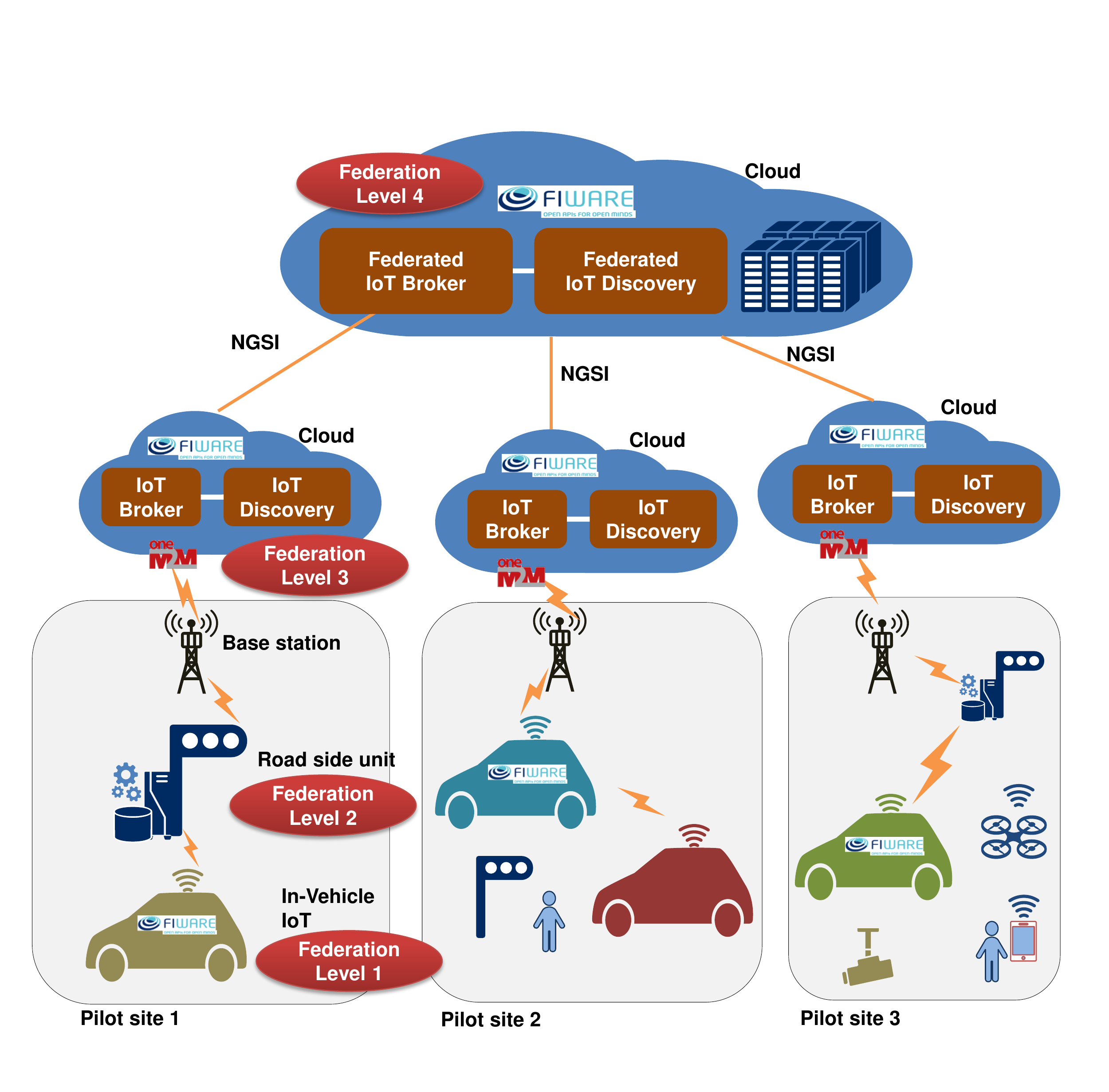}
\caption{Federation of large scale pilot sites through a FIWARE-based IoT platform.}
\label{Fig:AUTOPILOT_Federation}
\end{figure}

Together with common data models, the decentralized IoT architecture aims to enable long-distance automated travels, as described in~\cite{cirillo-sensors2019}. For instance, when an autonomous vehicle which is registered in Spain travels to the Netherlands, it should be able to operate using information coming from the city through which it is currently driving, while still pushing its key measurements to its home city. Information such as braking distance in different environments, which may indicate the status of its tires or the brake pads, may still be stored in the Spanish pilot site whereas the current location of the vehicle is stored in real-time on servers in the Netherlands. As another example, traffic situation information (e.g., congestion, traffic lights) from different cities are normally served by different vendors that do not federate data with each other. 
Enabling the sharing of data allows an autonomous car traveling from one country to another to learn the traffic situation in its current country. 

For realizing the federation of pilots, the combination of open-source FIWARE Aeron IoT Broker and NEConfMan IoT Discovery\footnote{\url{https://github.com/Aeronbroker}} components is used to transparently route IoT datasets and data streams from providers to consumers~\cite{Cirillo-LIoTS}. The transparency is given by the hidden brokering of the requests among IoT actors, where the IoT providers declare the available data and the IoT consumers specify the data of interest. An actor can play both roles, for instance, a car might be a data provider with geographic localization or a consumer of traffic information. The exchange might go from one pilot site to another as illustrated in Fig.~\ref{Fig:AUTOPILOT_Federation}. Federation starts from the vehicle (Level 1) with the in-vehicle IoT platform, where our lightweight version of the IoT Broker, called {\em ThinBroker} \footnote{\url{https://fogflow.readthedocs.io/en/stable/broker.html}}, can run on a server in the car. This component can run even in small devices such as a Raspberry Pi. 

The autonomous vehicles in the same region are connected (vehicle-to-vehicle communications) or as shown in pilot site 1, cars can connect to a roadside unit. In this case, RSUs are considered as the ``edges'' with processing and storage capabilities where IoT platform components are deployed as the Level 2 federation. In automotive scenarios, the IoT devices involved are not simply sensors but rather compound devices exposing many different interfaces. For that reason, instead of implementing several brand new IoT Agents, the existing interworking between oneM2M and other IoT technologies is leveraged~\cite{Kovacs2016}. As illustrated on the cloud side of Fig.~\ref{Fig:AUTOPILOT_Federation}, each pilot site has a oneM2M implementation for device-to-device communication as well as interworking between several IoT platforms. The cloud-side IoT platform is considered as Level 3 of the federation. This is formed again by IoT Brokers that (besides routing requests to/from oneM2M) also handle data locally which are managed by the pilot site administration.
Finally, there exists a federated IoT platform that operates across all pilot sites. Level 3 and 4 of the federation operate with common information models (data models). AUTOPILOT defines a new data model supporting the IoT information for autonomous driving vehicles, considering the AUTOPILOT use cases and the existing legacy models such as SENSORIS\footnote{\url{https://sensor-is.org/}} and DATEX II\footnote{\url{https://www.datex2.eu/}}. This data model covers IoT information such as vehicles, road infrastructure, road conditions, traffic situations (e.g., congestion), accidents, pedestrians, cyclists, vulnerable road user detection, events, road obstacles, potholes. The data is then mapped into the standard NGSI data format.

\subsection{Semantic Interoperability}

\begin{figure}
\centering
\includegraphics[width=0.9\columnwidth,trim={0cm 1cm 3cm 0cm},clip]{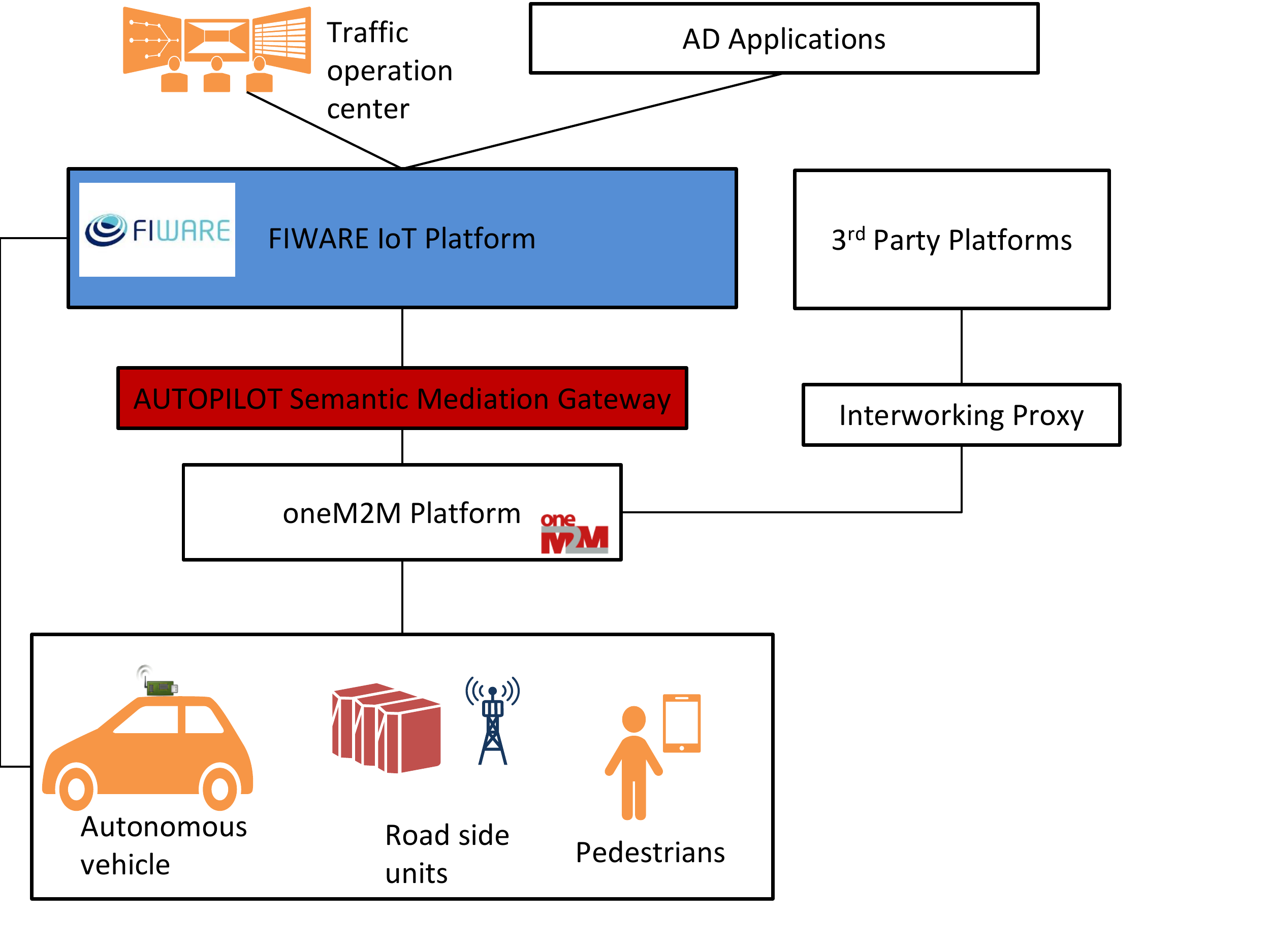}
\caption{Vehicle, RSUs, or pedestrians connecting to the FIWARE platform directly or through the oneM2M platform.}
\label{Fig:AUTOPILOTDeployment}
\end{figure}

Interworking between oneM2M and NGSI is achieved through semantic annotations~\cite{Kovacs2016}. In AUTOPILOT the interoperability features are provided by the Semantic Mediation Gateway (SMG) technology. SMG offers bidirectional context translations between oneM2M's {\em Mca} reference point and the NGSI interface. While the syntactic representations in oneM2M and NGSI are different, interoperability is enabled through the agreement on semantics, i.e. the same meaning, mapping homologous underlying concepts, which enables an SMG to do the translation.
Figure~\ref{Fig:AUTOPILOTDeployment} shows the basic setup for AUTOPILOT where the bottom layer consists of autonomous vehicles, RSUs, and pedestrians (e.g., people with smartphones) interacting with the top layer consisting of autonomous driving applications or traffic operation centers (top of Fig.~\ref{Fig:AUTOPILOTDeployment}). 
The interaction is provided through the FIWARE-based IoT platform. 
Some devices such as mobile devices of pedestrians connect and push data to the oneM2M platform. In some cases data is directly pushed to the FIWARE platform, or, otherwise, acquired by the SMG from oneM2M.
Nevertheless, other third-party platforms or components can receive information directly through oneM2M's Mca interface. 

The real-time system, consisting of in-vehicle components, oneM2M and FIWARE IoT platform, is tested at a major pilot site in the Netherlands. For instance, data coming from WiFi scanners of the autonomous vehicle and RSUs are used by analytics modules to estimate crowdedness information in the regions of interest. The contextualized crowdedness information is provided using the standardized data models in NGSI format. This information is then shared with the automated driving application to decide the travel route of the autonomous vehicles. As a result of having an agreement on common semantic concepts, the sharing of information is possible across all levels of the federation.


\subsection{The experience of IoT-augmented automated driving}

The AUTOPILOT project showcases that through a standard-based interworking approach different IoT platforms can run in harmony without causing significant extra costs or overhead. The main benefits of the proposed interworking approach can be summarized as follows.
\begin{itemize}
    \item Although there are many different and site-specific solutions, automated driving can be successful at the European scale, considering long travels.
    \item Different pilot sites do not have to adapt their specific devices or software to connect to different IoT platforms.
    \item Without much effort to adapt, autonomous vehicles can leverage a combination of various IoT services supported by different IoT platforms (e.g., crowd estimation service using FIWARE IoT platform, geofencing service by the Huawei OceanConnect platform).
    \item The federation at different layers can be achieved considering multiple pilot sites across Europe. 
\end{itemize}
One drawback of the interworking approach is to guarantee the adaptation of heterogeneous data from a vast number of possible services. To prevent service-specific efforts for conversion between different formats, common and standardized information models are a must. Hence, the AUTOPILOT has a specialized activity group having members from different stakeholders such as IoT architecture developers, pilot site leaders and use case leaders to agree on the common data models.

\section{The Road Ahead}
\label{sec:roadmap}

FIWARE has a very lively and active community that is continuously expanding. In this section, we include two key areas where FIWARE has the potential to improve.

\subsection{Semantics: NGSI-LD}

NGSI-LD is the evolution of the NGSI Context Interfaces~\cite{OMA-NGSI-Context} that has originally been standardized by the Open Mobile Alliance (OMA) and further developed in FIWARE. NGSI-LD is standardized as a group specification by the ETSI Industry Specification Group for cross-cutting Context Information Management (ISG CIM). The specification has been published~\cite{ETSI-CIM} in early 2019 and NGSI-LD is expected to become the new core interface for FIWARE GEs in the course of 2019\footnote{\url{https://github.com/ScorpioBroker/ScorpioBroker}}$^,$\footnote{\url{https://djane.io}}$^,$\footnote{\url{https://github.com/FIWARE/context.Orion-LD}}.

The main new feature is that NGSI-LD is now based on JSON-LD, which enables a semantic grounding. The \emph{LD} stands for \emph{linked data} which practically means that all elements are represented as URIs. Thus, the relevant concepts such as entity types can be explicitly defined in an ontology. Thus, the agreement between different applications and sources is on the level of explicitly specified semantics as provided in an ontology. This enables supporting a level of semantic interoperability on top of an otherwise heterogeneous IoT landscape, which is especially relevant in cross-cutting and large-scale application areas as can be found in smart cities. With the semantic modeling, existing semantic tools for ontology and rule-based reasoning can be used on information retrieved through many NGSI-LD requests. 

The new NGSI-LD  specification also foresees native support to historical data. It permits applications to directly request complete time series from the data context management together with complex IoT data query capabilities. 

\subsection{Privacy: Data Usage Control}

Soon, many advancements are expected in the field of distributed privacy and data usage control for securely sharing local domain data spaces. As presented in the previous sections, these features are a serious blocking point for moving from experimental IoT to a real global IoT. In the latter, sensitive data (e.g. healthcare data, industrial IoT data, crowdsourced data) share the same infrastructure with the less sensitive data. Topics such as continuous control on the usage of the data, also after data have been shared, shall be addressed. The collaboration between FIWARE and International Data Spaces Association (IDSA) is a strategic alliance since the requirements and reference architecture of the latter are the inputs for the evolution of the framework in the hands of the former. The initial results are already available~\cite{FIWARE4IDS}, showing the feasibility and the compatibility of the visions of FIWARE and IDSA.

\section{Acknowledgements}
This work has been partially funded by the European Union's Horizon 2020 Programme under Grant Agreement No. 731993 AUTOPILOT (AUTOmated driving Progressed by Internet Of Things), No. 723076 CPaaS.io (City Platform as a Service - integrated and open), and No. 732240 SynchroniCity (Delivering an IoT enabled Digital Single Market for Europe and Beyond). The content of this paper does not reflect the official opinion of the European Union. Responsibility for the information and views expressed therein lies entirely with the authors.

\ifCLASSOPTIONcaptionsoff
  \newpage
\fi

\bibliographystyle{IEEEtran}
{
\bibliography{Bib}
}

\end{document}